\documentclass[a4paper,aps,prd,10pt,preprintnumbers,showpacs,twocolumn,superscriptaddress,nofootinbib,amsmath,amssymb,gensymb,amsfonts]{revtex4-1}
\usepackage{graphicx}
\usepackage{cmap}
\usepackage[utf8]{inputenc}
\usepackage[T1]{fontenc}
\usepackage{nicefrac}
\usepackage{bm,amsmath,amssymb,amsfonts,gensymb,bbold}
\usepackage[colorlinks=true,linktocpage=true,linkcolor=blue,citecolor=blue,allcolors=blue]{hyperref}
\usepackage{amsmath}
\usepackage{makecell}
\usepackage{soul}
\usepackage{placeins}
\usepackage{epstopdf}
\graphicspath{{../}}

\begin{document}
\title{Non-Minimal Einstein–Yang–Mills Black Holes:\\ Fundamental Quasinormal Mode and Grey-Body factors versus Outburst of Overtones}
\author{B. C. Lütfüoğlu}
\email{bekir.lutfuoglu@uhk.cz}
\affiliation{Department of Physics, Faculty of Science, University of Hradec Kralove, \\
Rokitanskeho 62/26, Hradec Kralove, 500 03, Czech Republic. }

\begin{abstract}
Recently, an exact black hole solution in non-minimal Einstein–Yang–Mills theory was obtained, and its quasinormal modes were subsequently analyzed using the JWKB approximation \cite{Gogoi:2024vcx}. However, we demonstrate that this analysis lacks sufficient accuracy when studying modes with $\ell \leq n$, where $\ell$ is the multipole number and $n$ is the overtone number. To address this, we compute the quasinormal frequencies using the precise Leaver method. Our results show that while the fundamental mode deviates only slightly from the Schwarzschild value, the first few overtones exhibit significantly larger deviations, with the discrepancy growing rapidly with the overtone number. Moreover, beginning with the second overtone, we observe a striking phenomenon: the real part of the frequency tends to zero quickly as the non-minimal coupling constant increases. This combination of spectral stability in the fundamental mode and high sensitivity of the overtones suggests that the Yang–Mills contribution primarily deforms the metric in the near-horizon region, while the geometry quickly transitions back to a Schwarzschild-like form at larger radii. In addition, we compute the grey-body factors and confirm that they represent a more stable characteristic of the geometry, exhibiting a correspondence with quasinormal modes already at the first multipole. The Yang–Mills coupling enhances the grey-body factors, further increasing the transmission probability.
    
\end{abstract}
\maketitle

\section{Introduction}

The detection of gravitational waves from merging black holes by the LIGO and Virgo collaborations \cite{ligo1,ligo2,ligo3,ligo4,ligo5} has opened a new window into the strong-field regime of gravity. In particular, the ringdown phase of the signal, characterized by quasinormal modes (QNMs) \cite{review1,review2,review3,review4}, offers a powerful probe of the near-horizon geometry and provides a direct way to test general relativity in the dynamical, high-curvature regime. Simultaneously, advancements in high-resolution imaging of black holes, most notably by the Event Horizon Telescope (EHT) \cite{EventHorizonTelescope:2019dse,EventHorizonTelescope:2019ggy,Goddi:2016qax}, and the detection of X-ray and radio emissions from accretion disks have enriched our understanding of black hole environments through electromagnetic observations \cite{Bambi:2015kza}. The synergy between gravitational and electromagnetic channels, often referred to as multi-messenger black hole spectroscopy, holds the promise of constraining black hole parameters with unprecedented precision and testing the underlying spacetime geometry beyond the weak-field approximation.

Recent work \cite{Konoplya:2022pbc} challenges the widespread assumption that QNMs, particularly in the early ringdown phase, are largely insensitive to the near-horizon geometry of black holes. Using a general parametrization framework that preserves post-Newtonian asymptotics, it was shown that while the fundamental mode exhibits only modest deviations, the first few overtones display strong sensitivity to even small deformations localized near the event horizon. This so-called ``outburst of overtones'' implies that the early-time ringdown signal, in which the overtones are necessary \cite{Giesler:2019uxc,Giesler:2024hcr}, may carry detailed imprints of near-horizon modifications. The study further suggests that this overtone behavior represents a more robust observational signature than echoes \cite{Konoplya:2022pbc}, which typically arise at much later times when the gravitational wave signal has already significantly diminished \cite{Cardoso:2016rao}. These findings offer a promising avenue for probing quantum or modified gravity effects near the black hole horizon using future high-precision detectors such as LISA \cite{LISA:2022kgy}.

As a result, the behavior of overtones for black holes in various modified geometries—arising from quantum corrections, extra-dimensional models, or other extensions of general relativity—has recently been the subject of extensive investigation \cite{Konoplya:2025hgp,Konoplya:2024lch,Konoplya:2023ahd,Konoplya:2023ppx,Zinhailo:2024kbq,Zinhailo:2024jzt,Bolokhov:2023bwm,Stuchlik:2025ezz,Zhang:2024nny,Stashko:2024wuq}.

One such extended theory of gravity, the non-minimal Einstein–Yang–Mills (EYM) theory, admits black hole solutions \cite{Balakin:2006gv,Balakin:2015gpq} that practically coincide with the Schwarzschild geometry at large distances, but deviate significantly from it in the near-horizon region. The QNMs of a massless scalar field in this background have recently been studied using the Jeffreys-Wentzel–Kramers–Brillouin (JWKB) method in \cite{Gogoi:2024vcx}. However, that analysis is limited to the zeroth overtone ($n = 0$) with multipole numbers $\ell \geq 1$, and does not include the fundamental mode with $\ell = n = 0$. Moreover, the JWKB approximation is known to be asymptotic in nature and, even at high orders, does not guarantee reliable accuracy, particularly for low multipole numbers or overtones. 

In this work, we compute the quasinormal frequencies of black holes in non-minimal EYM gravity using the precise Leaver method \cite{Leaver:1985ax,Leaver:1986gd}. Our results show that while the fundamental mode experiences only a slight shift when the coupling parameter $\xi$ is introduced, the overtones undergo significant modifications. In particular, at the lowest multipole number $\ell=0$, the real part of the frequency for the second overtone rapidly tends to zero as $\xi$ increases, indicating a qualitative change in the spectrum. Furthermore, we demonstrate that the JWKB approach used in \cite{Gogoi:2024vcx} suffers from insufficient accuracy when studying higher overtones or the lowest multipole case $\ell=0$.

While QNMs are key spectral signatures of classical radiation around black holes, the grey-body factors encode crucial information about quantum, Hawking, radiation. They determine the fraction of Hawking radiation that escapes to infinity, modifying the black-body spectrum into the observable emission. Including grey-body factors in the analysis is essential not only for modeling realistic black hole evaporation, but also for probing the geometry near the horizon. In particular, grey-body factors are often more stable under geometric deformations than the quasinormal spectrum, making them a complementary diagnostic tool in the study of modified gravity, quantum-corrected spacetimes, or higher-dimensional theories. Their sensitivity to the structure of the effective potential further enables cross-validation of quasinormal mode results and provides an independent means of extracting physical information from black hole spacetimes.

The paper is organized as follows. In Sec.~II, we briefly review the non-minimal EYM black hole solutions and the corresponding wave equation for scalar perturbations. Sec.~III outlines the Leaver method and its implementation in our context, including the treatment of convergence issues and high-overtone stability. In Sec.~IV, we describe the sixth-order JWKB method with Padé approximants used for cross-checking our results. Sec.~V presents our main numerical findings for QNMs and grey-body factors, including both JWKB and quasinormal-mode-based estimates of transmission coefficients. Finally, Sec.~VI offers our conclusions and discussion of the implications of overtone behavior and grey-body factor stability in probing near-horizon structure.

\section{Non-Minimal Einstein–Yang–Mills Black Holes and the wave equation}

\begin{figure}
\resizebox{\linewidth}{!}{\includegraphics{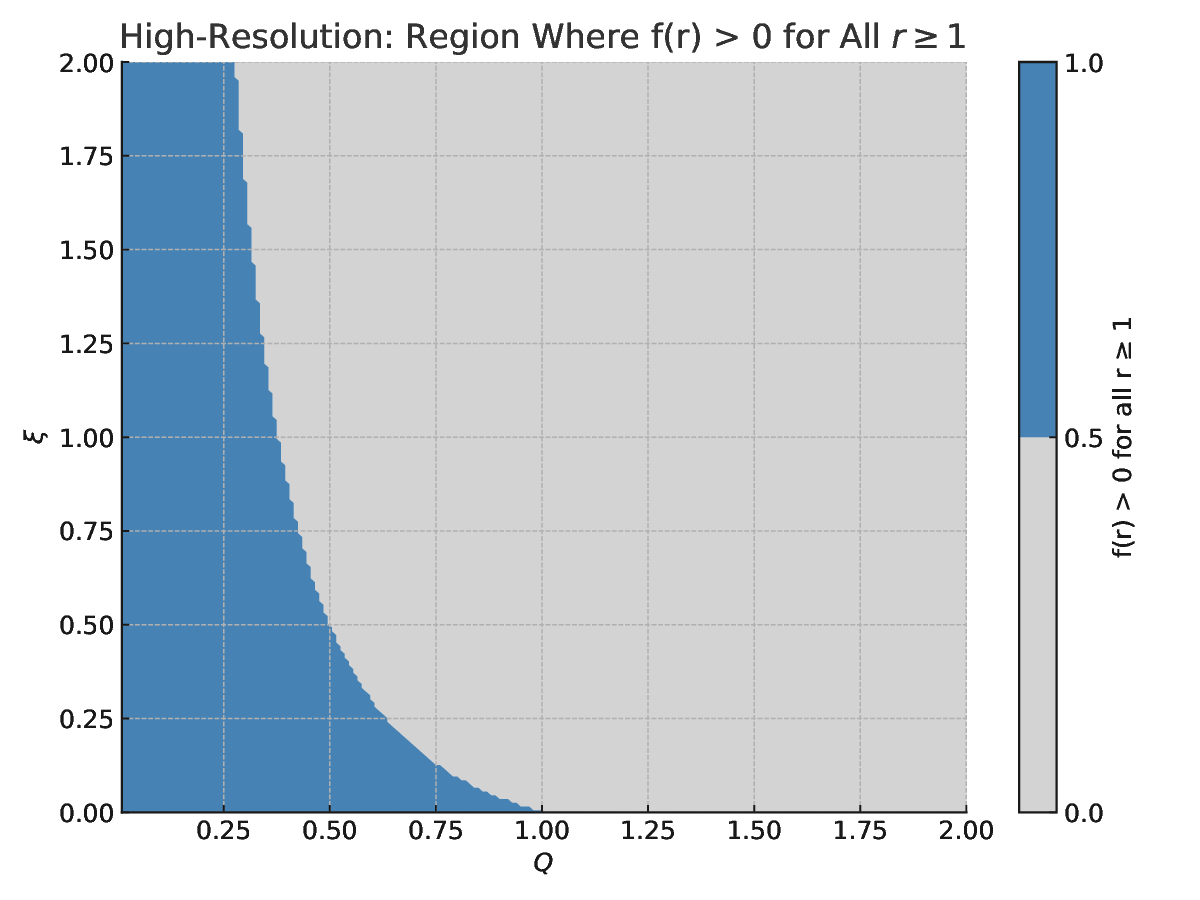}}
\caption{The parametric range where the black hole exist, once $r_{h}=1$.}\label{fig:BH}
\end{figure}

\begin{figure*}
\resizebox{\linewidth}{!}{\includegraphics{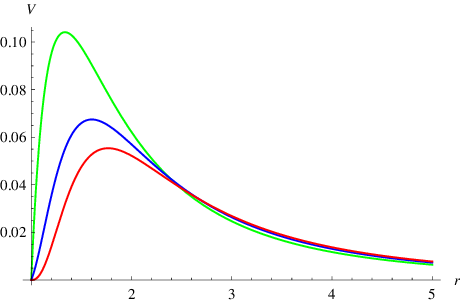}~~\includegraphics{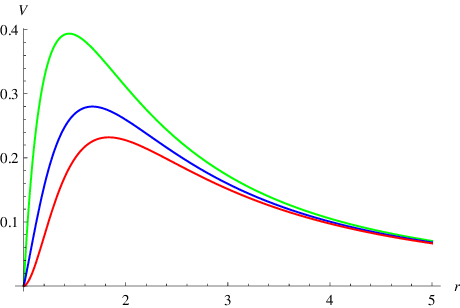}}
\caption{Effective potential for $\ell=0$ (left) and $\ell=1$ (right) perturbations $Q=0.1$, $\xi=0$ (green), $\xi=10$ (blue) and $\xi =16$ (red), $r_{h}=1$.}\label{fig:Potentials}
\end{figure*}

Non-minimal couplings between gravitational and gauge fields arise naturally in various extensions of general relativity, including semiclassical and string-inspired models. These couplings often reflect quantum corrections or effective descriptions of underlying high-energy physics. One interesting example is the EYM theory with non-minimal curvature coupling, which modifies how the spacetime geometry interacts with the Yang–Mills field.

In this context, we consider a theory defined by the following action:
\begin{equation}
\mathcal{S} = \int d^4x \sqrt{-g} \left[ \frac{R}{16\pi G} - \frac{1}{4} F^a_{\mu\nu} F^{a\mu\nu} + \frac{\xi}{4} R F^a_{\mu\nu} F^{a\mu\nu} \right],
\end{equation}
where \( R \) is the Ricci scalar, \( F^a_{\mu\nu} \) is the Yang–Mills field strength tensor associated with the \( SU(2) \) gauge group, and \( \xi \) is the non-minimal coupling constant. The last term in the action introduces a direct interaction between spacetime curvature and the gauge field, leading to non-trivial modifications of black hole solutions.

This theory admits static, spherically symmetric black hole solutions sourced by a Wu–Yang magnetic monopole configuration. The spacetime metric has the form:
\begin{equation}
ds^2 = -f(r)\,dt^2 + \frac{dr^2}{f(r)} + r^2(d\theta^2 + \sin^2\theta\,d\phi^2),
\end{equation}
where the metric function is given by:
\begin{equation}
f(r) = 1 + \left( \frac{r^4}{r^4 + 2\xi Q^2} \right) \left( \frac{Q^2}{r^2} - \frac{2M}{r} \right).
\end{equation}
Here, \( M \) is the mass of the black hole, \( Q \) is the magnetic charge of the Yang–Mills field, and \( \xi \) governs the strength of the non-minimal coupling. When \( \xi = 0 \), the standard minimally coupled EYM solution is recovered, and if \( Q = 0 \), the solution reduces to the Schwarzschild metric. The parametric range that permits the existence of an event horizon, expressed in units of the event horizon radius, is shown in Fig. \ref{fig:BH}.

The metric is asymptotically flat, and its behavior remains regular outside the event horizon provided the parameters lie within the physically allowed range. At large distances, \( f(r) \rightarrow 1 \), which ensures that conventional asymptotic boundary conditions for black hole perturbations can be applied.

To investigate the dynamical response of such spacetimes, one typically studies the evolution of a test field. Here, we consider a scalar field governed by the covariant Klein–Gordon equation:
\begin{equation}
\frac{1}{\sqrt{-g}} \partial_\mu \left( \sqrt{-g}\,g^{\mu\nu} \partial_\nu \Phi \right) = 0,
\end{equation}
where \( \mu \) is the mass of the scalar field. Using a standard decomposition,
\[
\Phi(t, r, \theta, \phi) = e^{-i \omega t} \frac{\Psi(r)}{r} Y_{\ell m}(\theta, \phi),
\]
one finds that the radial function \( \Psi(r) \) satisfies a Schrödinger-like wave equation:
\begin{equation}\label{waveEquation}
\frac{d^2\Psi}{dr_*^2} + \left( \omega^2 - V(r) \right)\Psi = 0,
\end{equation}
with the tortoise coordinate defined via \( dr_*/dr = 1/f(r) \). The effective potential takes the form:
\begin{equation}
V(r) = f(r) \left[ \frac{\ell(\ell+1)}{r^2} + \frac{f'(r)}{r}\right].
\end{equation}
This potential is positive-definite outside the event horizon and therefore, supports only damped quasinormal oscillations (see Fig. \ref{fig:Potentials}).

\section{The Leaver method}

For asymptotically flat spacetimes, the QNM condition requires purely ingoing waves at the event horizon and purely outgoing waves at infinity, expressed as:
\[
\Psi(r) \sim 
\begin{cases}
e^{-i \omega r_*}, & r \to r_h \quad (\text{horizon}), \\
e^{+i \omega r_*}, & r \to \infty,
\end{cases}
\]
where \( r_* \) is the tortoise coordinate and \( \omega \) is the complex QNM frequency. The real part of $\omega$ determines the real oscillation frequency, while the imaginary part is proportional to its damping rate.

The Leaver method is a robust semi-analytical technique for calculating QNMs of black holes. It was originally developed for Schwarzschild and Kerr spacetimes but can be generalized to a broad class of spherically symmetric spacetimes with suitable boundary behavior. The method reduces the wave equation to a recurrence relation for series coefficients, and quasinormal frequencies are extracted from the condition of convergence of the resulting continued fraction.

The solution $\Psi(r)$ is expressed as a Frobenius-type series expansion around the event horizon:
\begin{equation}
    \Psi(r) = e^{i \omega r_*} \left(\frac{r - r_h}{r}\right)^\alpha \sum_{n=0}^\infty a_n \left(\frac{r - r_h}{r}\right)^n,
\end{equation}
where $\alpha$ ensures ingoing behavior at the horizon and the exponential factor accounts for asymptotic behavior.

Substituting into the wave equation, one obtains a recurrence relation of the general form:
\begin{equation}
    \sum_{j=0}^{k} A_{n,j} a_{n-j} = 0,
\end{equation}
where $k$ is the order of the recurrence relation (e.g., $k=2$ for a three-term recurrence).

For many black hole spacetimes, especially with suitable coordinate redefinitions, the recurrence simplifies to a three-term relation:
\begin{equation}
    \alpha_n a_{n+1} + \beta_n a_n + \gamma_n a_{n-1} = 0 \quad (n \geq 1),
\end{equation}
with initial condition:
\begin{equation}
    \alpha_0 a_1 + \beta_0 a_0 = 0.
\end{equation}

The convergence condition for the series yields a continued fraction equation:
\begin{equation}
    \frac{a_1}{a_0} = -\frac{\beta_0}{\alpha_0} = \frac{\gamma_1}{\beta_1 -} \ \frac{\alpha_1 \gamma_2}{\beta_2 -} \ \frac{\alpha_2 \gamma_3}{\beta_3 -} \cdots 
\end{equation}
Solving this continued fraction numerically gives the quasinormal frequencies $\omega$.


In some black hole spacetimes or for certain fields (e.g., non-minimally coupled, massive, or with nontrivial asymptotics), the recurrence relation may involve more than three terms:
\begin{equation}
    \sum_{j=0}^{k} A_{n,j} a_{n-j} = 0, \quad k > 2.
\end{equation}

Such relations can be reduced to a three-term form using Gaussian elimination. For instance, a four-term recurrence:
\begin{equation}
    a_{n+1} + b_n a_n + c_n a_{n-1} + d_n a_{n-2} = 0,
\end{equation}
can be transformed into an equivalent three-term relation via recursive elimination of higher-order terms. The resulting three-term relation can then be treated with the standard continued fraction technique.

The Leaver method is highly accurate and converges rapidly. It is particularly well-suited for black holes with smooth effective potentials and analytic behavior at the horizon and infinity. It has been successfully applied to Schwarzschild, Reissner--Nordstr\"om, Kerr, Kerr--Newman, and numerous modified gravity black holes.

\begin{figure*}
\resizebox{\linewidth}{!}{\includegraphics{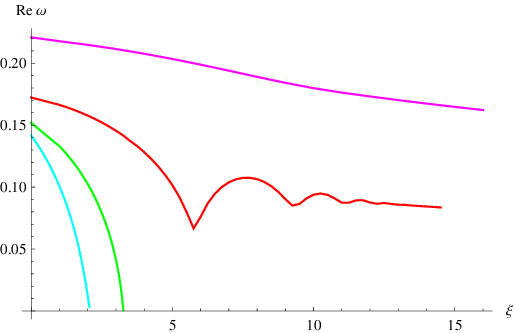}\includegraphics{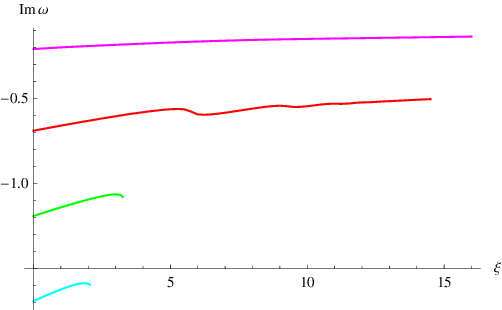}}
\caption{Fundamental quasinormal mode and the first three overtones for $\ell=0$ perturbations as a function of $\xi$; $Q=0.1$, $r_{h}=1$.}\label{fig:QNM1}
\end{figure*}
\begin{figure*}
\resizebox{\linewidth}{!}{\includegraphics{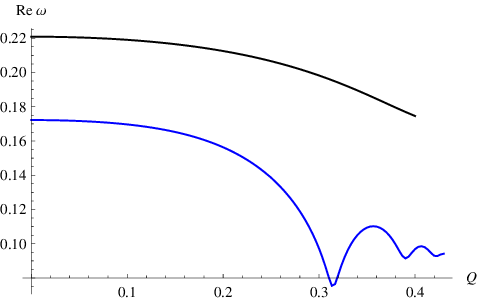}\includegraphics{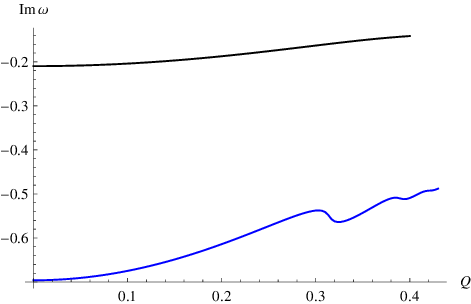}}
\caption{Fundamental quasinormal mode for $\ell=0$ perturbations as a function of $Q$; $\xi=0.5$ (black) and $\xi =0.7$ (blue); $r_{h}=1$.}\label{fig:QNM3}
\end{figure*}
\begin{figure*}
\resizebox{\linewidth}{!}{\includegraphics{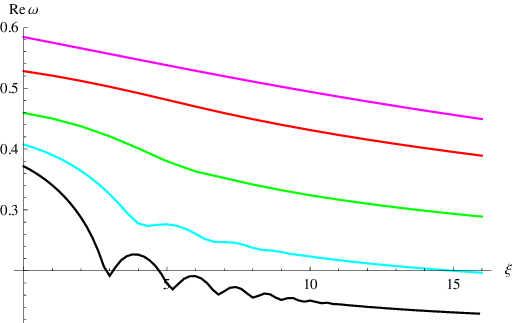}\includegraphics{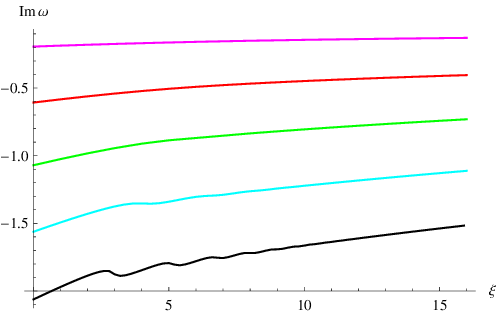}}
\caption{Fundamental quasinormal mode and the first four overtones for $\ell=1$ perturbations as a function of $\xi$; $Q=0.1$, $r_{h}=1$.}\label{fig:QNM2}
\end{figure*}

\begin{figure*}
\resizebox{\linewidth}{!}{\includegraphics{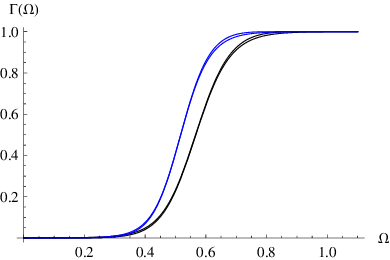}\includegraphics{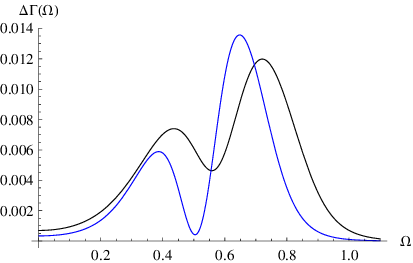}}
\caption{Grey-body factors as a function of real frequency $\Omega$: $\xi=0.1$, $Q=0.3$ (black) and  $\xi=0.7$, $Q=0.3$ (blue) calculated via the higher order JWKB approach and via the correspondence with the QNMs together with the difference between the results obtained by the two methods; $\ell=1$, $r_{h}=1$.}\label{fig:GBF1}
\end{figure*}

\begin{figure*}
\resizebox{\linewidth}{!}{\includegraphics{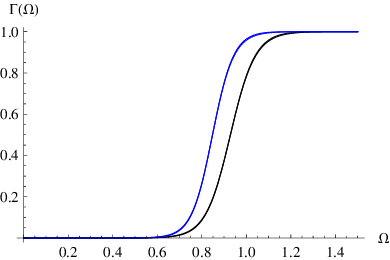}\includegraphics{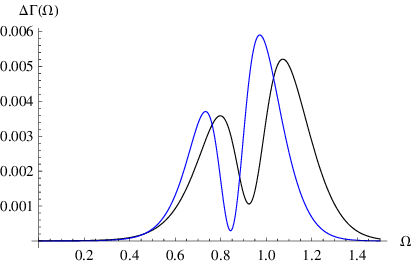}}
\caption{Grey-body factors as a function of real frequency $\Omega$: $\xi=0.1$, $Q=0.3$ (black) and  $\xi=0.7$, $Q=0.3$ (blue) calculated via the higher order JWKB approach and via the correspondence with the QNMs together with the difference between the results obtained by the two methods; $\ell=2$, $r_{h}=1$.}\label{fig:GBF2}
\end{figure*}

\begin{table*}
\centering
\begin{tabular}{|c|c|c|c|c|c|c|c|}
\hline
\hline
$\ell$ & Schwarzschild & JWKB & Leaver & Effect-Re (\%) & Effect-Im (\%) & Error-Re (\%) & Error-Im (\%) \\
\hline
0 & 0.220910 - 0.209791 i & 0.215463 - 0.188247 i & 0.212169 - 0.187323 i & -3.956815 & -10.709706 & 1.552536 & 0.493266 \\
1 & 0.585872 - 0.195320 i & 0.560171 - 0.175740 i & 0.560056 - 0.175724 i & -4.406423 & -10.032767 & 0.020534 & 0.009105 \\
2 & 0.967288 - 0.193518 i & 0.924228 - 0.174443 i & 0.924215 - 0.174446 i & -4.452965 & -9.855414 & 0.001407 & -0.001720 \\
3 & 1.350732 - 0.192999 i & 1.290473 - 0.174102 i & 1.290471 - 0.174103 i & -4.461359 & -9.790724 & 0.000155 & -0.000574 \\
\hline
\hline
\end{tabular}
\caption{Comparison of fundamental QNM frequencies with $\ell = 0$, $1$, $2$ for the Schwarzschild limit ($Q=\xi=0$),  and Yang-Mills BH ($Q=0.3$, $\xi =0.1$) found by the 6th order JWKB approximation with Padé approximants ($\tilde{m} =\tilde{n}=3$), and precise Leaver method. ``Effect'' quantifies the relative deviation between Schwarzschild and Yang-Mills QNM frequencies, while ``Error'' reflects the discrepancy between JWKB and Leaver results.}
\label{table:qnm-comparison}
\end{table*}
\begin{table*}
\centering
\begin{tabular}{|c|c|c|c|c|c|c|c|}
\hline
\hline
$\ell$ & Schwarzschild & JWKB & Leaver & Effect-Re (\%) & Effect-Im (\%) & Error-Re (\%) & Error-Im (\%) \\
\hline
0 & 0.172234 - 0.696105 i & 0.079284 - 0.482935 i & 0.085351 - 0.486338 i & -50.444744 & -30.134391 & -7.108294 & -0.699719 \\
1 & 0.528897 - 0.612515 i & 0.394414 - 0.400637 i & 0.394686 - 0.400797 i & -25.375640 & -34.565243 & - 0.068916 & -0.040095 \\
2 & 0.927701 - 0.591208 i & 0.714145 - 0.389644 i & 0.714120 - 0.389635 i & -23.022612 & -34.095107 & 0.003501 & 0.002310 \\
3 & 1.321343 - 0.584570 i & 1.023224 - 0.386826 i & 1.023211 - 0.386827 i & -22.562802 & -33.827087 & 0.001271 & -0.000259 \\
\hline
\hline
\end{tabular}
\caption{Comparison of the first overtone with $\ell = 0$, $1$, $2$ for the Schwarzschild limit ($Q=\xi=0$),  and Yang-Mills BH ($Q=0.3$, $\xi =1.5$) found by the 6th order JWKB approximation with Padé approximants ($\tilde{m} =\tilde{n}=3$), and precise Leaver method. ``Effect'' quantifies the relative deviation between Schwarzschild and Yang-Mills QNM frequencies, while ``Error'' reflects the discrepancy between JWKB and Leaver results.}
\label{table:qnm-comparison2}
\end{table*}
\begin{table*}
\centering
\begin{tabular}{|c|c|c|c|c|c|c|c|}
\hline
\hline
$n$ & Schwarzschild & JWKB & Leaver & Effect-Re (\%) & Effect-Im (\%) & Error-Re (\%) & Error-Im (\%) \\
\hline
0 & 0.220910 - 0.209791 i & 0.223732 - 0.209400 i & 0.220021 - 0.207219 i & -0.402426 & -1.225982 & 1.686657 & 1.052510 \\
1 & 0.172234 - 0.696105 i & 0.177302 - 0.713622 i & 0.171827 - 0.686728 i & -0.236306 & -1.347067 & 3.186344 & 3.916252 \\
2 & 0.151484 - 1.202157 i & 0.157499 - 1.227848 i & 0.150370 - 1.185789 i & -0.735391 & -1.361553 & 4.740972 & 3.546921 \\
3 & 0.140820 - 1.707355 i & 0.154860 - 1.729000 i & 0.138339 - 1.684140 i & -1.761824 & -1.359717 & 11.942402 & 2.663686 \\
\hline
\hline
\end{tabular}
\caption{Comparison of quasinormal frequencies for $\ell = 0$ and overtones $n = 0$ to $3$ in the Schwarzschild limit ($Q = \xi = 0$) and Yang–Mills BH ($Q = 0.1$, $\xi = 0.1$), calculated via sixth-order JWKB approximation and the Leaver method. ``Effect'' denotes the relative difference between Schwarzschild and Leaver values, while ``Error'' quantifies the discrepancy between JWKB and Leaver.}
\label{table:qnm-overtones}
\end{table*}

To overcome convergence limitations in Leaver’s continued fraction method due to additional regular singularities inside the unit circle after the usual variable transformation, the solution is first expanded as a power series near the horizon, then analytically continued to a midpoint within the convergence radius \cite{Rostworowski:2006bp}. At this intermediate point—chosen so that the irregular singularity at infinity becomes the nearest—one applies the continued fraction condition to extract quasinormal frequencies. This modified prescription ensures that the method remains robust even when extra singularities obstruct direct application of Leaver’s original formulation.

To enhance numerical stability and convergence of the continued fraction, especially for high overtone modes, we employ the Nollert improvement \cite{Nollert:1993zz}. This involves replacing the asymptotic tail of the continued fraction with an analytic approximation to the minimal solution ratio, significantly accelerating convergence and improving accuracy in the determination of quasinormal frequencies.

The Leaver method together with the integration through the mid point and Nollert improvement has been used in great number of works (see, for instance \cite{Konoplya:2004uk,Zinhailo:2024jzt,Rosa:2011my,Ohashi:2004wr,Konoplya:2017tvu,Bolokhov:2024bke,Bolokhov:2023bwm,Bolokhov:2023ruj,Kanti:2006ua,Konoplya:2007zx,Bhagwat:2019dtm,Berti:2003zu,Zhidenko:2006rs}), demonstrating efficiency and precision not only for the dominant modes, but also for high overtones.

\section{JWKB Method with Padé Approximants}

The JWKB approximation has long been a standard semi-analytic tool for calculating QNMs of black holes in asymptotically flat and de Sitter spacetimes, where the effective potential takes the form of a single potential barrier. Originally developed to third order by Schutz and Will, and extended to sixth order by Iyer and Will, the JWKB method was later pushed to thirteenth order with recursive expressions for higher-order correction terms \cite{Iyer:1986np,Konoplya:2003ii,Matyjasek:2017psv}.

The general idea of the method is to treat the wave equation for perturbations given by Eq. \ref{waveEquation},
as a one-dimensional Schrödinger-like equation, where \(r_*\) is the tortoise coordinate and \(V(r)\) is the effective potential. The JWKB approach yields an analytic approximation for the complex frequencies \(\omega\) by expanding around the peak of the potential barrier. At the \(N\)th order, the JWKB formula takes the form:
\begin{equation}
i \frac{\omega^2 - V_0}{\sqrt{-2 V_0''}} - \sum_{j=2}^{N} \Lambda_j = n + \frac{1}{2}, \quad n = 0,1,2,\dots,
\end{equation}
where \(V_0\) is the height of the effective potential, \(V_0''\) its second derivative with respect to the tortoise coordinate, and \(\Lambda_j\) are higher-order correction terms.

However, despite its analytical appeal, the JWKB series is asymptotic and does not necessarily converge as \(N\) increases. In fact, it may start to diverge beyond a certain order, especially for low multipole numbers \(\ell\) or high overtone numbers \(n\). To improve the convergence and accuracy of the method, Padé approximants have been introduced \cite{Matyjasek:2017psv}. These are rational functions constructed from the JWKB series which often provide better numerical performance and faster convergence than the raw polynomial expansion.

In the Padé-improved JWKB method, the JWKB series for $\omega$ is replaced by its \(\text{Padé}_{\tilde{n}}^{\tilde{m}}\) approximant, typically with \(\tilde{m} = \tilde{n} = N/2\), although other combinations are also used. This modification substantially enhances the accuracy of the quasinormal mode estimates, particularly for low overtones and in cases where the potential is not sharply peaked.

The JWKB method with Padé approximants is especially effective for test fields and for gravitational perturbations in spacetimes where the effective potential has a single, well-defined peak and decays monotonically at both infinities \cite{Konoplya:2001ji,Zinhailo:2019rwd,Xiong:2021cth,Skvortsova:2024atk,Skvortsova:2023zca,Skvortsova:2024wly,Hamil:2024nrv,Liu:2024wch,Konoplya:2005sy,Kodama:2009bf,Malik:2025ava,Malik:2024tuf,Malik:2024elk,Churilova:2021tgn,Dubinsky:2024hmn,Lutfuoglu:2025hwh,Lutfuoglu:2025hjy,Gogoi:2023kjt,Yang:2022ifo}.  However, it should be used cautiously when the potential develops multiple extrema or deep negative gaps, such as in some higher-curvature or quantum-corrected black hole spacetimes, where more robust numerical methods like time-domain integration or the continued fraction method are required for reliable results.

In \cite{Gogoi:2024vcx}, the authors applied the sixth-order JWKB method with Padé approximation to compute the QNMs of massless scalar fields in this background. While this technique is useful in the high-frequency (eikonal) regime, it is known to produce substantial errors when applied to fundamental modes (\(n = 0\)) with low multipole numbers (\(\ell \leq n\)) \cite{Konoplya:2019hlu}. In these cases, the numerical error can exceed the very physical effects under investigation — such as small shifts in the frequencies caused by the Yang–Mills charge or the non-minimal coupling.  In the following section, we will show that this is the case for the nonminimal EYM theory.

\vspace{5mm}

\section{Quasinormal modes and grey-body factors}

First of all, we discuss the case
of the lowest multipole $\ell = 0$. Within the Leaver method it is convenient to use units of the
radius of the event horizon $r_{h} = 1$. Then the mass is parametrized as follows:
\[
M = \frac{1}{2} \left( 1 + Q^2 + 2\xi Q^2 \right).
\]
The range of parameters allowing for the black hole event horizon is shown in Fig. \ref{fig:BH}.

The QNMs for various parameter values are presented in Figs. \ref{fig:QNM1}-\ref{fig:QNM2}. In Fig.~\ref{fig:QNM1}, we observe that while the real part of the fundamental mode ($\ell = n = 0$) changes only moderately — by about 10–30\% as $\xi$ increases — the first overtone already decreases by more than a factor of two. The second and higher overtones behave quite differently, rapidly approaching zero. A similar phenomenon has been reported in~\cite{Konoplya:2024lch} for certain quantum-corrected black holes and in~\cite{Zinhailo:2024jzt} for massive fields around brane-localized black holes. The deviation in the damping rate from the Schwarzschild limit also grows with the overtone number, although at a relatively moderate pace. In  Fig.~\ref{fig:QNM3}, we also see that the charge affects the quasinormal frequencies, resulting in a decrease of both the real part and the damping rate. However, at the near extreme $\xi$ the change of the frequency becomes non-monotonic.

For $\ell = 1$ perturbations, whose spectrum is shown in Fig.~\ref{fig:QNM2}, we observe that the overtones deviate from their Schwarzschild limits at a slower rate compared to the $\ell = 0$ case. Nevertheless, the fourth overtone already shows a clear deviation — exceeding a factor of two relative to its Schwarzschild counterpart. We conclude that the overtone outburst occurs for all multipoles, but its strength increases as the multipole number decreases.

In the regime of high multipole numbers, $\ell \gg 1$, an analytic expression for the QNMs can be derived by performing a double expansion in both inverse powers of $\kappa = \ell + 1/2$ and the Yang–Mills charge $Q$. The position of the maximum of the effective potential admits the following expansion:
\begin{widetext}
\begin{equation}
r_{max} = 3 M-\frac{2 Q^2 \left(27 M^2+7 \xi \right)}{81 M^3}-\frac{4
   Q^4 \left(729 M^4+783 M^2 \xi +163 \xi ^2\right)}{19683
   M^7} + O(Q^5, \kappa^{-1}).
\end{equation}
Using the above expression for the location of the maximum into the first-order JWKB formula and expanding the result in powers of $\kappa^{-1}$, we obtain the expression for quasinormal modes
$$\omega =\frac{\kappa }{3 \sqrt{3} M}+\frac{Q^2 \kappa  \left(27
   M^2+4 \xi \right)}{486 \sqrt{3} M^5}+\frac{Q^4 \kappa 
   \left(9477 M^4+4752 M^2 \xi +640 \xi ^2\right)}{472392
   \sqrt{3} M^9}$$
\begin{equation}
   -\frac{i K}{3 \sqrt{3}
   M}-\frac{i Q^2 K \left(27 M^2-44
   \xi \right)}{1458 \sqrt{3} M^5}+\frac{i Q^4 K \left(2187
   M^4+18576 M^2 \xi +4672 \xi ^2\right)}{472392 \sqrt{3}
   M^9}+O\left(Q^5, \kappa^{-1}\right),
\end{equation}
\end{widetext}
where $K = n +(1/2)$. When $\xi = Q = 0$, the above relations go over into the well-known eukonal expressions for the Schwarzchild black hole \cite{Mashhoon:1982im}.

In~\cite{Gogoi:2024vcx}, it was argued that in the eikonal regime, quasinormal frequencies respect the correspondence with the properties of unstable null geodesics—namely, the Lyapunov exponent and the angular frequency at the circular photon orbit~\cite{Cardoso:2008bp}—and hence, with the radius of the black hole shadow. Thus, 
\[
\omega_{QNM} = \Omega_c \ell - i \left( n + \frac{1}{2} \right) \lambda,
\]
where:
\[
\Omega_c = \left. \frac{\sqrt{f(r)}}{r} \right|_{r = r_c},
\]
and $\lambda$ is the Lyapunov exponent. 

However, an explicit analytic expression for the quasinormal frequencies in this regime was not derived in \cite{Gogoi:2024vcx}. In the present work, we confirm the validity of this correspondence for the considered black hole spacetime by explicitly deriving the eikonal formula. Despite the existence of several known counterexamples to the geodesic–QNM correspondence~\cite{Konoplya:2022gjp,Bolokhov:2023dxq,Konoplya:2017wot}, our analytic results support its applicability in this case. It is worth noting that while the correspondence between QNMs and null geodesics is not strictly guaranteed for test fields, it typically holds in practice due to the JWKB-suitable nature of the effective potential. However, this correspondence may break down for gravitational perturbations in EYM theory. In particular, there are known counterexamples where the effective potential deviates significantly from the standard centrifugal form $g_{tt} \ell (\ell+1)/r^2$, potentially leading to instabilities. Such deviations commonly arise in theories with higher curvature corrections~\cite{Takahashi:2011du,Dotti:2004sh,Konoplya:2017lhs,Konoplya:2020bxa,Takahashi:2011qda,Cuyubamba:2016cug,Chowdhury:2024uzd}.

This analytic expression could be extended beyond the eikonal approximation using the general approach of \cite{Konoplya:2023moy}, which has been recently used in \cite{Dubinsky:2024rvf,Malik:2024qsz,Malik:2024elk,Malik:2024tuf,Malik:2024voy}.

It is also worth noting that the JWKB method used in~\cite{Gogoi:2024vcx} is insufficient for analyzing quasinormal frequencies with $\ell \leq n$, as the physical effect becomes comparable in magnitude to the numerical error—particularly when the Yang–Mills correction is relatively small. This limitation is evident from Tables~\ref{table:qnm-comparison}–\ref{table:qnm-overtones}.

While QNMs are characteristics that are sensitive to the near horizon deformations of the background spacetime, the grey-body factors are shown to be much more stable function \cite{Oshita:2023cjz,Oshita:2024fzf,Wu:2024ldo,Dubinsky:2024nzo}. 

The grey-body factor quantifies the probability that a wave emitted near a black hole horizon escapes to infinity, accounting for the potential barrier surrounding the black hole. Unlike an ideal black body, a black hole does not radiate perfectly; part of the radiation is scattered back by the curvature-induced effective potential. Mathematically, the grey-body factor is defined through the wave equation with scattering boundary conditions: purely ingoing waves at the event horizon and a combination of incoming and outgoing waves at spatial infinity. For a wave function $\psi(r)$ governed by a Schrödinger-like equation with an effective potential $V(r)$, the boundary conditions read $\psi \sim e^{-i\Omega x}$ near the horizon ($x \to -\infty$) and $\psi \sim A_{\text{out}} e^{i\Omega x} + A_{\text{in}} e^{-i\Omega x}$ at infinity ($x \to \infty$), where $x$ is the tortoise coordinate. The grey-body factor is then given by the transmission coefficient $\Gamma = |T|^2 = 1 - |R|^2$, where $R = A_{\text{in}} / A_{\text{out}}$ is the reflection coefficient.

The transmission coefficient $|T_\ell|^2$, which determines the grey-body factor for each multipole number $\ell$, can be accurately computed using the JWKB approximation. A refined version of the method expresses the reflection coefficient as
\[
R = \left(1 + e^{-2i\pi K}\right)^{-1/2},
\]
where $K$ is a function of the frequency and the shape of the effective potential. It satisfies the transcendental equation
\[
K - i\frac{\Omega^2 - V_\text{max}}{\sqrt{-2 V_\text{max}''}} - \sum_{i=2}^6 \Lambda_i(K) = 0,
\]
with $V_\text{max}$ and $V_\text{max}''$ denoting the maximum of the potential and its second derivative, respectively, and $\Lambda_i$ representing higher-order JWKB corrections. This approach, which generalizes the standard first-order JWKB result, provides a reliable estimation of grey-body factors across a broad frequency range, especially for intermediate and high energies.

An alternative and efficient method for estimating grey-body factors is based on their correspondence with QNMs. Although these quantities are defined under different boundary conditions—grey-body factors from a scattering problem allowing for incoming radiation from infinity, and QNMs under purely outgoing boundary conditions—they are closely related in the eikonal regime. In this high-frequency limit, it has been shown that the transmission coefficient can be approximated in terms of the fundamental quasinormal mode. Specifically, for a spherically symmetric black hole, the transmission coefficient takes the form
\[
\Gamma_\ell(\Omega)  = \left(1 + \exp\left[\frac{2\pi\left(\Omega^2 - \text{Re}(\omega_0)^2\right)}{4\, \text{Re}(\omega_0)\, |\text{Im}(\omega_0)|} \right]\right)^{-1} + \mathcal{O}(\ell^{-1}),
\]
where $\Omega$ is the real frequency in the scattering problem, and $\omega_0$ is the fundamental quasinormal mode frequency. This expression (Eq.~3.5 in~\cite{Konoplya:2024lch}) is exact in the eikonal limit and provides a remarkably simple link between the two spectral characteristics. Extensions beyond the eikonal regime are possible by incorporating overtone corrections, leading to highly accurate estimates of grey-body factors even at low multipole numbers \cite{Malik:2024cgb,Bolokhov:2024otn,Konoplya:2024vuj,Pedrotti:2025upg,Hamil:2025cms,Dubinsky:2024vbn,Skvortsova:2024msa}.

From Figs.~\ref{fig:GBF1} and \ref{fig:GBF2}, we observe that increasing the coupling constant $\xi$ leads to larger grey-body factors. This behavior can be readily explained by examining the effective potential in units of the event horizon radius: higher values of $\xi$ correspond to a lower potential barrier, as illustrated in Fig.~\ref{fig:Potentials}. As a result, a greater portion of the radiation is able to penetrate the barrier surrounding the black hole. We also find that the grey-body factors change only moderately with increasing charge or coupling $\xi$, confirming that they are indeed a more stable characteristic of black hole radiation.

\section{CONCLUSIONS}

In this work, we have revisited the quasinormal mode spectrum of black holes in non-minimal EYM theory. Using the precise Leaver method, we have demonstrated that while the fundamental mode is only moderately shifted by the non-minimal coupling, the overtones—especially at low multipole numbers—undergo substantial modifications. This outburst of overtones provides a sensitive probe of the near-horizon geometry and complements the relatively stable behavior of the fundamental mode. In the eikonal limit we find analytic expressions for QNMs and show that they respect the correspondence with the null geodesics. 

We have also calculated the grey-body factors of these black holes via two independent approaches: a high-order JWKB method and the recently established correspondence with QNMs. Both methods show excellent agreement in the eikonal regime, and our analysis confirms that grey-body factors remain more robust under deformations of the geometry, such as those introduced by the Yang–Mills coupling. This stability supports their role as a reliable diagnostic of the underlying black hole structure.

 Future work may extend this approach to perturbations of other types of massless fields, as well as to massive fields, which have a number of peculiar features in the frequency and time domains, such as arbitrarily long lived modes and oscillatory slowly decaying tails \cite{Ohashi:2004wr,Konoplya:2006gq,Koyama:2001qw,Koyama:2001ee,Churilova:2019qph,Konoplya:2017tvu,Zhang:2018jgj,Gibbons:2008gg,Gonzalez:2022upu}, as well as  potentially observational signatures in the Pulsar Timing Array experiment \cite{Konoplya:2023fmh,NANOGrav:2023gor}.

\vspace{4mm}
\begin{acknowledgments}
The author is grateful to Excellence Project PrF UHK 2205/2025-2026.
\end{acknowledgments}

\FloatBarrier
\bibliography{bibliography}
\end{document}